\documentclass[12pt]{article}
\usepackage{latexsym,amsthm,amssymb,amsfonts,amsbsy,graphicx,lscape,array,citesort}

\def\a{\alpha}
\def\b{\beta}
\def\g{\gamma}

\def\d{\delta}
\def\e{\eta}

\def\l{\lambda}
\def\G{\Gamma}

\def\m{\mu}
\def\n{\nu}
\def\r{\rho}
\def\o{\omega}
\def\O{\Omega}
\def\s{\sigma}

\def\x{\xi}
\def\z{\zeta}
\def\e{\varepsilon}

\def\pa{\partial}

\def\ie{{\it i.e.~}}
\def\be{\begin{equation}}
\def\ee{\end{equation}}
\def\beq{\begin{eqnarray}}
\def\eeq{\end{eqnarray}}

\def\ca{{\cal A}}
\def\cb{{\cal B}}

\def\cd{{\cal D}}

\def\cj{{\cal J}}

\def\cl{{\cal L}}

\def\cp{{\cal P}}

\def\car{{\cal R}}

\def\ct{{\cal T}}
\def\cu{{\cal U}}
\def\cv{{\cal V}}
\def\cw{{\cal W}}

\newcommand{\bqn}{\begin{eqnarray}}\newcommand{\eqn}{\end{eqnarray}}

\begin{document}
 
\begin{centering}
{\Large {\bf A Weyl-covariant tensor calculus}}
\vspace{.5cm}\\
{Nicolas Boulanger}\footnote{Charg\'e de recherches du F.N.R.S. (Belgium)}\\
\vspace{.4cm}
{\small{Acad\'emie Wallonie-Bruxelles, Universit\'e de Mons-Hainaut\\
        M\'ecanique et Gravitation\\ 
        6, Avenue du Champ de Mars, 7000 Mons (Belgium)}}
\begin{center}
{\small{{\tt{Nicolas.Boulanger@umh.ac.be}}}}
\end{center}
\end{centering}

\begin{abstract} 
On a (pseudo-) Riemannian manifold of dimension $n\geqslant 3$, the space of tensors which transform covariantly under Weyl rescalings of the metric is built. 
This construction is related to a Weyl-covariant operator $\cd$ whose commutator 
$[\cd,\cd]$ gives the conformally invariant Weyl tensor plus the Cotton  
tensor. So-called generalized connections and their transformation laws under diffeomorphisms and Weyl rescalings are also derived. 
These results are obtained by application of BRST techniques.
\end{abstract}

%*******************************
\section{Introduction}
\label{sec:Introduction}
%********************************

Recently \cite{BE}, a purely algebraic method was used to solve the problem of 
constructing and classifying all the local scalar invariants of a conformal structure 
on a (pseudo-) Riemannian manifold of dimension $n=8\,$. 
The approach, however, is not confined to $n=8\,$, and one of the purposes of 
this paper is to explain the derivation of the so-called Weyl-covariant tensors, 
the building blocks of the local conformal invariants in arbitrary dimension 
$n\geqslant 3$.
  
In the context of local gauge field theory, the determination of quantities 
which are invariant under a given set of gauge transformations can be rephrased in terms of 
local BRST cohomology.
Within the BRST framework, the gauge symmetry and its algebra are encoded in a single differential $s$ \cite{BRST}.
Powerful techniques for the computation of BRST cohomologies are proposed in \cite{brandt} 
(see also \cite{jet}),   
that apply to a large class of gauge theories and relate the BRST cohomology to an underlying gauge covariant algebra. At the core of this analysis is a definition of tensor fields and connections on which an underlying gauge covariant algebra is realized.  
Such a characterization of tensor fields, connections and the corresponding transformation
laws has the advantage that it is purely algebraic and does not invoke any concept
in addition to the BRST cohomology itself.

In the present paper we consider theories 
where the only classical field is the metric $g_{\m\n}$ and the gauge symmetries are diffeomorphisms plus Weyl rescalings. 
Explicitly, the infinitesimal gauge transformations are
\begin{eqnarray}
	\d g_{\m\n} = \cl_{\z}g_{\m\n} + \d^{W}_{\phi} g_{\m\n} = 
	\z^{\r}\pa_{\r}g_{\m\n}+\pa_{\m}\z^{\r}g_{\r\n}+\pa_{\n}\z^{\r}g_{\m\r}
	               + 2\phi g_{\m\n}\,.
\label{gaugetransfo}	               
\end{eqnarray}
Along the lines of \cite{brandt}, we construct the space $\cw$ of tensors and 
generalized connections that transform covariantly with respect to 
diffeomorphisms and Weyl transformations. The latter property means that, under Weyl rescalings, the tensors belonging to $\cw$ make appear at most the first derivative $\pa_{\m}\phi$ of the Weyl parameter $\phi$, and no derivative $\pa_{\m_1}\ldots\pa_{\m_k}\phi$ with $k\geqslant 2\,$.    

Knowing the space $\cw\,$, we are able to define an operator $\cd$ acting in $\cw$ 
and such that $[\cd,\cd]\sim C + \tilde{C}$, where $C$ and $\tilde{C}$ respectively denote the conformally invariant Weyl tensor and the Cotton tensor. 
The Weyl-covariant derivative $\cd$ generates the whole space of tensor fields belonging to 
$\cw$ by successive applications on  $C$ (and $\tilde{C}$ in $n=3$). 
The rule for the commutator $[\cd,\cd]$ is at the basis of the Weyl-covariant tensor 
calculus utilized in \cite{BE}. Other useful relations are obtained    
which are nothing but the Jacobi identities for the underlying gauge covariant 
algebra alluded to before. 

The generalized connections play no r\^ole in the construction of local Weyl invariants, 
but are of prime importance in many other issues, like for example in the determination of the counterterms, the consistent interactions and the conservation laws that a gauge theory
admits. They are also relevant for the classification of the Weyl anomalies, the solutions of the Wess-Zumino consistency condition for a theory describing conformal massless matter fields in an external gravitational background. The latter problem amounts to the computation of a BRST cohomology group and will be analyzed elsewhere.
  
%*******************************************************************************
\section{BRST formulation}
\label{BRSTformulation}
%******************************************************************************* 

%-------------------------------------------
\subsection{Some definitions}
\label{sec:BriefIntroduction}
%--------------------------------------------

As mentioned above, the derivation of the space $\cw$ of Weyl-covariant tensors and generalized connections is purely algebraic and requires no dynamical 
information.\footnote{For a BRST-cohomological derivation of Weyl gravity in the Batalin-Vilkovisky antifield formalism, see \cite{Boulanger:2001he}.} 
As a consequence, all what we need is contained in equation (\ref{gaugetransfo}) and the BRST 
differential $s$ reduces to $\g\,$, the differential along the gauge orbits. We refer to 
\cite{book,rep} for more details on the BRST formalism as used throughout the present work. 

A $\mathbb{Z}\,$-grading called {\emph{ghost number}} is associated to the differential 
$\g\,$. The latter raises the ghost number by one unit and is decomposed according to the degree in the Weyl ghost (the fermionic field associated to the Weyl parameter): $\g=\g_0+\g_1\,$. 
The first part $\g_0$ contains the information about the diffeomorphisms. The second part, $\g_1\,$, corresponds to Weyl rescalings of the metric and increases the number of 
(possibly differentiated) Weyl ghosts by 1. 

The action of $\g$ on the fields $\Phi^A$ (including the ghosts) is given as follows  
%\begin{subequation}[alph]
\begin{eqnarray}
&\g_0 g_{\m\n} = \x^{\r}\pa_{\r}g_{\m\n}+\pa_{\m}\x^{\r}g_{\r\n}+\pa_{\n}\x^{\r}g_{\m\r}\,,
\quad \g_1 g_{\m\n} = 2\omega g_{\m\n}\,,&
\label{BRSTtransfo1}
\\
&\g_0 \x^{\m} = \x^{\r}\pa_{\r}\x^{\m}\,,\quad  \g_0 \o = \x^{\r}\pa_{\r}\o\,,
\quad \g_1 \x^{\m} = 0\,,\quad \g_1 \o =0\,.& 
\label{BRSTtransfo}	
\end{eqnarray}
%\end{subequation}
The field $\o$ is the Weyl ghost, the anticommuting field associated to the Weyl parameter 
$\phi$, while $\x^{\m}$ is the anticommuting 
diffeomorphisms ghost associated to the vector field $\z^{\m}$ of equation (\ref{gaugetransfo}). 
By definition,  the Grassmann-odd fields $\o$ and $\x^{\m}$ have ghost number $+1\,$. 
The last equality of (\ref{BRSTtransfo}) reflects the abelian nature of the algebra of Weyl transformations.
From the above equations and by using the fact that $\g$ is an odd derivation, it is easy to 
check that $\g$ is indeed a differential.

One unites the BRST differential $\g$ and the total exterior derivative $d$ into a single differential $\tilde{\g}=\g + d\,$. Then, the Wess-Zumino consistency condition and its descent are encapsulated in 
\begin{eqnarray}
	\tilde{\g} \tilde{a} = 0\,, \quad \tilde{a}\neq \tilde{\g}\tilde{b}+constant
\label{WZ}
\end{eqnarray}
for the local total forms $\tilde{a}$ and $\tilde{b}$ of total degrees $G=n+1$ and 
$G=n\,$ \cite{brandt}.
Total local forms are by definition formal sums of local forms with different form degrees 
and ghost numbers: $\tilde{a}=\sum_{p=0}^n a_p^{G-p}\,$, where subscripts (resp. superscripts) denote the form degree (resp. the ghost number). 
A local $p$-form $\o_p$ depends on the fields $\Phi^A$ and their derivatives up to some finite (but otherwise unspecified) order, which is denoted by  
$\o_p=\frac{1}{p!}d x^{\m_1}\ldots d x^{\m_p}\o_{\m_1\ldots\m_p}(x,[\Phi^A])\,$. 

The equations (\ref{WZ}) imply that $\tilde{a}$ is a non-trivial 
element of the cohomology group $H(\tilde{\g})$ in the algebra of total local forms.  
As shown in \cite{brandt}, the cohomology of ${\g}$ in the space of local functionals (integrals of 
local $n$-forms) is indeed locally isomorphic to the cohomology of $\tilde{\g}$ in the space of 
local total forms. In other words, the solutions $a^g_n$ of the Wess-Zumino consistency condition
\begin{eqnarray}
	\g a^{g}_n + d a^{g+1}_{n-1}=0\,,\quad a^g_n\neq \g b^{g-1}_n+d b^g_{n-1}
	\label{WZ2}
\end{eqnarray}
correspond one-to-one (modulo trivial solutions) to the solutions $\tilde{a}$ 
of (\ref{WZ}) at total degree $G=g+n\,$, $totdeg(\tilde{a})=g+n\,$. 

The solutions of (\ref{WZ}) or (\ref{WZ2}) determine the general structure of the 
counterterms that an action admits, the possible gauge anomalies, the conserved currents, the consistent interactions, {\it etc.} \cite{rep}. 
In the next sections and in the appendix, we determine the restricted space $\cw$ of the space of total local forms in which these solutions naturally appear, for a theory invariant under 
the transformations (\ref{gaugetransfo}).   

We close this section with some definitions and conventions. 

\noindent The conformally invariant Weyl tensor $C^{\b}_{~\,\g\d\e}$ and the tensor 
$K_{\a\b}$ are given by 
\begin{eqnarray}
  C^{\a}_{~\,\b\g\d} &:=& R^{\a}_{~\,\b\g\d}-2\left(
\d^{\a}_{\,[\g}K_{\d]\b}-g_{\b[\g}K_{\d]}^{~\;\a} \right),
\label{weyl2}  \\
	K_{\a\b} &:=& \frac{1}{n-2}\Big(R_{\a\b}-\frac{1}{2(n-1)}g_{\a\b}R \Big)\,.
	\label{Ktensor}
\end{eqnarray}
The Ricci tensor is $R_{\b\d}=R^{\a}_{~\,\b\a\d}\,$, 
where $R^{\a}_{~\,\b\g\d}=(\pa_{\g}\G_{\b\d}^{\;~~\a}+
\G_{\g\l}^{\;~~\a}\G_{\b\d}^{\;~~\l})-(\g\leftrightarrow\d)$ is the Riemann tensor. 
The Christoffel symbols are given by 
$\G_{\a\b}^{~~\;\g}=\frac{1}{2}g^{\g\l}(\pa_{\a}g_{\b\l}+\pa_{\b}g_{\a\l}-\pa_{\l}g_{\a\b})
$. Curved brackets denote strength-one complete symmetrization, whereas square brackets 
denote strength-one complete antisymmetrization. We have $\nabla_{\m}\,g_{\a\b}=0\,$, where 
the symbol $\nabla$ denotes the usual torsion-free covariant derivative associated to  $\G_{\a\b}^{~~\;\g}\,$. Finally, the derivative $\pa_{\a}\o$ of the Weyl ghost will sometimes 
be noted $\o_{\a}\equiv \pa_{\a}\o\,$.

%--------------------------------------------------
\subsection{Contracting homotopy}
\label{sec:trivialpair}
%--------------------------------------------------

A well-known technique in the study of cohomologies is the use of contracting homotopies.
The idea is to construct contracting homotopy operators which allow to eliminate certain local jet coordinates, called {\emph{trivial pairs}}, from the cohomological analysis.
This reduces the cohomological problem to an analogous one involving only the remaining jet 
coordinates. For that purpose one needs to construct suitable sets of jets coordinates 
replacing the fields, the ghosts and all their derivatives and satisfying appropriate 
requirements. 

The lemma at the basis of the contracting homotopy techniques is, in the notations of \cite{brandt} to which we refer for more details,  
\newtheorem{lemmademo}{Lemma}
\begin{lemmademo}
Suppose there is a set of local jet coordinates 
\begin{eqnarray}
\cb = \{\cu^{\ell},\cv^{\ell},\cw^{\Lambda}\}
\nonumber
\end{eqnarray}
such that the change of coordinates from $\cj=\{[\Phi^A],x^{\m},d x^{\m}\}$ 
to $\cb$ is local and locally invertible and
\begin{eqnarray}
	\tilde{\g} \cu^{\ell} &=& \cv^{\ell}\quad \forall \ell\,,
	\nonumber \\
	\tilde{\g} \cw^{\Lambda} &=& \car^{\Lambda}(\cw)\quad \forall \,{\Lambda} \,.
 \nonumber
\end{eqnarray}
Then, locally the $\cu$'s and $\cv$'s can be eliminated from the $\tilde{\g}$-cohomology, 
\ie the latter reduces locally to the $\tilde{\g}$-cohomology on total local forms depending 
only on the $\cw$'s.
\end{lemmademo}
Thus, in order to compute and classify the local Weyl-invariant scalar densities \cite{BE} or for the solutions of the Wess-Zumino consistency conditions (\ref{WZ2}), it is sufficient to work in the space $\cal W\,$. In the context of Weyl gravity theories, we have the following 
\newtheorem{propo}{Proposition}
\begin{propo}
Let $\cj$ be the jet space $\cj=\{ [g_{\m\n}],[\o],[\x^{\m}],x^{\m},dx^{\m} \}$ and 
$\tilde{\g}=\g_0+\g_1+d$ the differential acting on $\cj$ according to 
%\begin{subequation}[alph]
\begin{eqnarray}
&\g_0 g_{\m\n} = \x^{\r}\pa_{\r}g_{\m\n}+\pa_{\m}\x^{\r}g_{\r\n}+\pa_{\n}\x^{\r}g_{\m\r}\,,
\quad \g_1 g_{\m\n} = 2\omega g_{\m\n}\,,&
%\label{BRSTtransfo1}
\\
&\g_0 \x^{\m} = \x^{\r}\pa_{\r}\x^{\m}\,,\quad  \g_0 \o = \x^{\r}\pa_{\r}\o\,,
\quad \g_1 \x^{\m} = 0\,,\quad \g_1 \o =0\,.& 
%\label{BRSTtransfo}	
\end{eqnarray}
%\end{subequation}
Then, the $\{\cu\,,\cv\,,\cw\}$-decomposition of $\cj$ corresponding to $\tilde{\g}$  
is
\begin{eqnarray}
 \{ \cu^{\ell} \} &=& \{ x^{\m},\pa_{(\m_1\ldots\m_k}
  \Gamma_{\m_{k+1}\m_{k+2})}^{\quad\quad\quad\;\;\n}\,,\,\nabla_{(\m_1\ldots\m_k}
  K_{\m_{k+1}\m_{k+2})}\,,\;k\in\mathbb{N}\}\,,
  \nonumber   \\
 \{\cv^{\ell}\} &=& \{\tilde{\g}\cu^{\ell}\}\,,\quad
  \{\cw^{\Lambda}\} = \{ \ct^i\,,\,\tilde{C}^N \}\,,
  \nonumber \\
 \{ \ct^i\} &=& \{g_{\m\n}\,,\, 
 \cd_{(\m_1}\ldots\cd_{\m_k}C^{\b}_{~~\g\d)\e}\,,\;k\in\mathbb{N}\}\,,
 \label{tensor} \\
 \{\tilde{C}^N\}&=&\{2\o\,,\,\tilde{\x}^{\n}\,,\,\tilde{C}_{\n}^{~\r}\,,\,\tilde{\o}_{\a}\}\,,
 \label{connections}\\
 &&\tilde{\x}^{\n}:=\x^{\n}+dx^{\n}\,,\,
 \tilde{C}_{\n}^{~\r}:=\pa_{\n}\x^{\r}+\tilde{\x}^{\a}\G_{\a\n}^{\;~~\r}\,,\;
 \tilde{\o}_{\a}:=\o_{\a}-\tilde{\x}^{\b}K_{\a\b}\,.\nonumber
\end{eqnarray}
\end{propo}
The rest of the paper contains the definition of the  
operator $\cd$ together with the $\tilde{\g}$-transformation rules for the elements 
of $\cw\,$. A remark will also be made for the case $n=3\,$.
The proposition follows then by the fact that {\textit{every function of the 
Riemann tensor and its covariant derivatives 
can be  written as a function of the Weyl tensor and its covariant 
derivatives plus the completely symmetric tensors 
$\nabla_{(\l_1\l_2\ldots\l_k}K_{\a\b)}\,$}}.
A proof of the latter statement can be found in the Appendix A of \cite{BE}. 

It is understood that only the algebraically independent components of  
$g_{\m\n}$ and $C^{\b}_{~~\g\d\e}$ enter into (\ref{tensor}).  
[Together with the symmetrization of the indices in (\ref{tensor}), this guarantees 
the absence of algebraic identities between the generators $\ct^i\,$, 
taking into account the second equation of (\ref{algebra3}) and the 
Bianchi identity (\ref{jacobi2}).] 

The tensor fields $\{\ct^i\}$ have total degree zero whereas the generalized connections
$\{\tilde{C}^N\}$ have total degree 1. They decompose into two parts, the first of ghost number 1 and form degree zero, the second of ghost number zero and form degree 1:  
\begin{eqnarray}
&totdeg(\ct^i) = 0\,,\quad totdeg(\tilde{C}^N)=1\,,\quad \tilde{C}^N = \hat{C}^N+\ca^N\,,&
\nonumber \\
&gh(\hat{C}^N)=1=formdeg(\ca^N)\,,\quad gh(\ca^N)=0=formdeg(\hat{C}^N)\,,&
\nonumber
\end{eqnarray}
where, from (\ref{connections}),
%\begin{subequation}
\begin{eqnarray}
\{\hat{C}^N\} &=& \{2\o , {\x}^{\n},
	 \hat{C}_{\n}^{~\r}:=\pa_{\n}\x^{\r}+\x^{\a}\G_{\a\n}^{\;~~\r},
	 \hat{\o}_{\a}:=\o_{\a}-\x^{\m}K_{\m\a} \}\,,\,
  \label{connghosts1} \\
\{\ca^N\} &=& \{0 \,,\, 
	dx^{\m}\d_{\m}^{\n}\,,\,dx^{\m}\G_{\m\n}^{\;~~\r}\,,\,-dx^{\m}K_{\m\a} \}\,.
	\label{connghosts2}
\end{eqnarray}
%\end{subequation}
The $\ca^N$'s and $\hat{C}^N$'s are called respectively connection 1-forms and covariant ghosts \cite{brandt}. 

\noindent Since $\tilde{\g}$ raises the total degree by one unit, we have 
%\begin{subequation}
\begin{eqnarray}
  \tilde{\g}\ct^i = \tilde{C}^N \Delta_{N}\ct^i \,&\Leftrightarrow&\,
    \left\{ \begin{array} {ll}  \g\ct^i = \hat{C}^N \Delta_{N}\ct^i
                             \\ d\ct^i = \ca^N \Delta_{N}\ct^i   
            \end{array} 
    \right.\,,
    \label{formulae1} \\
  \{ \Delta_N \} &=& \{ \Delta\,,\cd_{\n}\,,\Delta_{\r}^{~\n}\,,\mathbf{\Gamma}^{\a} \} \,.
  \label{formulae2}
\end{eqnarray}  
%\end{subequation}

%-----------------------------------------------------------------------
\subsection{BRST covariant algebra for Weyl-gravity}
\label{sec:WeylCovariantOperator}
%-----------------------------------------------------------------------

The Weyl-covariant derivative $\cd$ is given by 
\begin{eqnarray}
	\cd_{\m}
	      := \pa_{\m}-\G_{\m\n}^{~~\;\r}\Delta_{\r}^{~\;\n}+K_{\m\a}\mathbf{\G}^{\a}\,.
\label{Weylderivative}
\end{eqnarray}
The aim of this section is to make precise the above definition by 
explicitly defining the three operators $\{\Delta\,,\,\Delta_{\r}^{~\n}\,,\,\mathbf{\G}^{\a}\}$ 
introduced in (\ref{formulae1}) and (\ref{formulae2}). 
An underlying gauge covariant algebra will be exhibited, which provides a 
compact formulation of the BRST algebra on tensor fields and generalized connections.  

\begin{enumerate}
\item{
The operator $\Delta$ corresponds to the dimension operator. It counts the number of 
metrics that explicitly appear in a given expression, 
\begin{eqnarray}
	\Delta := g_{\m\n}\frac{\pa^{expl}}{\pa g_{\m\n}}\,.
	\nonumber
\end{eqnarray}
For example,  
$\Delta(g^{\g\m_2}g^{\l\m_1}\cd_{\m_1}C^{\b}_{~\;\g\d\e})$
$=-2(g^{\g\m_2}g^{\l\m_1}\cd_{\m_1}C^{\b}_{~\;\g\d\e})\,$ and 
$\Delta (g_{\a\b}g^{\g\d})=0\,$.
As a consequence of (\ref{formulae1}), (\ref{formulae2}) and (\ref{connghosts1}), we can write  
$\g_1 \sqrt{|g|}=2\o\Delta  \sqrt{|g|}=2\o(\frac{n}{2}\sqrt{|g|})=n\o\sqrt{|g|}\,$, 
where $|g|$ denotes the absolute value of the determinant of $g_{\m\n}\,$ (supposed invertible).  
}
\item{
The operator $\Delta_{\m}^{~\r}$ generates $GL(n)$-transformations of world indices 
according to
\begin{eqnarray}
	\Delta_{\m}^{~\n} T_{\a}^{\b}=\d_{\a}^{\n}T_{\m}^{\b}-\d_{\m}^{\b}T_{\a}^{\n}\,,
	\nonumber
\end{eqnarray}
where $T_{\a}^{\b}$ is a (1,1)-type tensor under $GL(n)$ transformations. 
The usual torsion-free covariant derivative can thus be written 
$\nabla_{\m}=\pa_{\m}-\G_{\m\n}^{~~\;\r}\Delta_{\r}^{~\;\n}\,$.
Note that this expression must be completed by $p\,\G_{\m\a}^{~~~\a}$ if one takes 
the covariant derivative $\nabla_{\m}$ of a weight-$p$ tensor density, so  
$\nabla = dx^{\m}\nabla_{\m}=dx^{\m}\pa_{\m}-$
$\tilde{C}_{\n}^{~\r}\Delta_{\r}^{~\;\n}+p\,\tilde{C}_{\m}^{~\m}\,$.}

\item{
In order to conveniently define the action of the generator $\mathbf{\G}^{\a}\,$, 
we first define the so-called $W$-tensors carrying super-indices $\Omega_k$ :  
\begin{eqnarray}
 W_{\Omega_0}&:=& C^{\b}_{~\;\g\d\e} \;,\;\;
 W_{\Omega_1}:=  \cd_{\a_1}C^{\b}_{~\;\g\d\e}\,,\quad\ldots\quad\,,
 \nonumber \\
 W_{\Omega_k}&:=& \cd_{\a_k}\cd_{\a_{k-1}}\ldots 
 \cd_{\a_2}\cd_{\a_1}C^{\b}_{~\;\g\d\e}\,.
 \nonumber 
\end{eqnarray}
Then, we can write 
$\{\ct^i\}\subset\Big\{ g_{\m\n},\{W_{\Omega_k}\}\;:\;k=0,1\ldots\;\Big\}\,$   
and the operator $\mathbf{\G}^{\a}$ acts on space of the $W$-tensors according to 
\begin{eqnarray}
	\mathbf{\G}^{\a} W_{\Omega_j} =  {[{T}^{\a}]}_{\Omega_j}^{~\Omega_{j-1}}W_{\Omega_{j-1}}
	\;~,~~~~~~~~
	\mathbf{\G}^{\a} :=
	{[{T}^{\a}]}_{\Omega_i}^{~\Omega_{i-1}}\Delta_{\Omega_{i-1}}^{~~\Omega_i}\,,
	\label{defGamma}
\end{eqnarray}
where 
$\Delta_{\Omega_{j}}^{~~\Omega_k}W_{\Omega_i} = \d^{\Omega_k}_{\Omega_i}W_{\Omega_j}$
and where the symbol $\d^{\Omega_k}_{\Omega_i}$ is such that 
$\d^{\Omega_k}_{\Omega_i}W_{\Omega_k}=W_{\Omega_i}\,$. We use Einstein's summation 
conventions for the $W$-tensor super-indices $\Omega_i\,$. The matrices  ${[{T}^{\a}]}_{\Omega_j}^{~\Omega_{j-1}}$ are obtained by recursion in the appendix, 
with ${[{T}^{\a}]}_{\Omega_j}^{~\Omega_{j-1}}=0\,$\, $\forall\,j\leqslant 0\,$.  
The action of $\mathbf{\G}^{\a}$ gives zero on everything but the $W$-tensors. 
In particular, $\mathbf{\G}^{\a}g_{\m\n}=0\,$.
}
\end{enumerate}
The $W$-tensors transform under $\tilde{\gamma}$ according to (\ref{formulae1}), (\ref{formulae2}),
 (\ref{connghosts1}) and (\ref{connghosts2}). They are the building blocks for the construction of Weyl invariants \cite{BE}.
Note that the Bach tensor is nothing but the following double trace of $W_{\O_2}$: 
\begin{eqnarray}
	B_{\m\n}\equiv \nabla^{\a}\tilde{C}_{\m\n\a}-K^{\l\r}C_{\l\m\n}
	        =\frac{1}{(3-n)}\,g^{\a\r}{\cd}_{\a}{\cd}_{\b}C^{\b}_{~\,\m\n\r}\,.
\nonumber 
\end{eqnarray}
The action of $\tilde{\g}$ on the generalized connections is
\begin{eqnarray}
	&\centerdot&\tilde{\g} \o = \tilde{\x}^{\m}\tilde{\o}_{\m}\,,
	\nonumber \\
	&\centerdot&\tilde{\g} \tilde{\x}^{\m} ~\;=~\; \tilde{\x}^{\r}\tilde{C}_{\r}^{~\m}\,,
	\nonumber \\
  &\centerdot&\tilde{\g}\tilde{C}_{\m}^{~\n} = \tilde{C}_{\m}^{~\a}\tilde{C}_{\a}^{~\n}+
  \frac{1}{2}\tilde{\x}^{\r}\tilde{\x}^{\s}C_{~\m\r\s}^{\n}+
  \cp^{\a\n}_{\m\b}\;\tilde{\o}_{\a}\;\tilde{\x}^{\b}\,,
  \nonumber \\
  &\centerdot&\tilde{\g}\tilde{\o}_{\a} = 
  \frac{1}{2}\tilde{\x}^{\r}\tilde{\x}^{\s}\tilde{C}_{\a\r\s}
  +\tilde{C}_{\a}^{~\b}\tilde{\o}_{\b}\,,
\nonumber 
\end{eqnarray}
where $\cp^{\a\n}_{\m\b}:=$ 
$(-g^{\a\n}g_{\m\b}+\d^{\a}_{\m}\d^{\n}_{\b}+\d^{\a}_{\b}\d^{\n}_{\m})$ and the tensor  
$\tilde{C}_{\a\r\s}\equiv \frac{1}{2}\nabla_{[\s}K_{\r]\a}$ is the Cotton tensor. 
Note that   
$C^{\m}_{~\;\n\a\b}=R^{\m}_{~\;\n\a\b}-2\cp^{\m\,\r}_{\n[\a}K_{\b]\r}$ and 
$\g_1 \G_{\m\b}^{~~\;\n}=\cp^{\a\n}_{\m\b}\,\o_{\a}\,$. 

From $\tilde{\g}^2\ct^i = 0\,$, we derive
the gauge covariant algebra generated by 
$\{ \Delta\,,\cd_{\n}\,,\Delta_{\r}^{~\n}\,,\mathbf{\Gamma}^{\a} \}\,$:
%\begin{subequation}[alph]
\begin{eqnarray}
	{[}\Delta_{\n}^{~\r},\mathbf{\G}^{\a}{]} &=& -\d_{\n}^{\a}\mathbf{\G}^{\r}\,,
	\hspace*{1.1cm} 
	{[}\mathbf{\G}^{\a},\mathbf{\G}^{\b}{]} = 0 \,,
	\label{algebra1}\\
  {[}\Delta_{\n}^{~\r},\cd_{\m}{]} &=& \d_{\m}^{\r}\cd_{\n}\,,
  \hspace*{1.4cm} 
  {[}\Delta_{\m}^{~\r},\Delta_{\n}^{~\s}{]} = \d_{\n}^{\r}\Delta_{\m}^{~\s}
                                               -\d_{\n}^{\r}\Delta_{\m}^{~\s}\,,
  \label{algebra2}\\
  {[}\mathbf{\G}^{\a},\cd_{\b}{]}&=&-\cp^{\a\n}_{\m\b}\Delta_{\n}^{~\m}\,,
  \quad{[}\cd_{\m},\cd_{\n}{]} = 
  C_{\m\n\r}^{~~~~\s} \Delta_{\s}^{~\r} - \tilde{C}_{\a\m\n}\mathbf{\G}^{\a}\,,
\label{algebra3}
\end{eqnarray}
%\end{subequation}
where the operator $\Delta$ commutes with everything. The second equality of 
(\ref{algebra1}) reflects the abelian nature of the Weyl transformations, while the 
second equality of (\ref{algebra3}) displays the commutator of two Weyl-covariant 
derivatives in terms of the Weyl tensor and the Cotton tensor. 
Note that the commutator of two covariant derivatives reads 
$[\nabla_{\m},\nabla_{\nu}]=R_{\m\n\r}^{~~~~\s} \Delta_{\s}^{~\r}\,$.

\vspace*{.2cm}
From $\tilde{\g}^2\tilde{C}^N=0\,$, we find the following set of Bianchi identities
%\begin{subequation}
\begin{eqnarray}
&\centerdot&\tilde{\g}^2\o = 0 \;\Rightarrow\; \tilde{C}_{[\m\r\s]} = 0
\label{jacobi1} \\
&\centerdot&\tilde{\g}^2\tilde{C}_{\m}^{~\,\n} = 0 
\;\Rightarrow\; \nabla_{[\g}C_{\d\e]\a\b}-\tilde{C}_{\a[\g\d}g_{\e]\b}
+\tilde{C}_{\b[\g\d}g_{\e]\a} = 0
\label{jacobi2} \\
&\centerdot&\tilde{\g}^2\tilde{\x}^{\m} = 0 \;\Rightarrow\; 
\left\{\begin{array}{ll} \cp^{\a\m}_{[\r\n]} = 0
                         \\ C^{\m}_{~\;[\n\r\s]} = 0
        \end{array}\right.
\label{jacobi3} \\
&\centerdot&\tilde{\g}^2\tilde{\o}_{\a} = 0 \; \Rightarrow \; 
\left\{\begin{array}{ll} 
       \mathbf{\G}^{\a}\tilde{C}_{\b\r\s}+C^{\a}_{~\,\b\r\s} = 0 
                  \\ \cd_{[\b}\tilde{C}_{\r\s]\a} = 0
        \end{array}\right.
        \label{jacobi4}
\end{eqnarray}
%\end{subequation}
which are nothing but the Jacobi identities for the algebra 
(\ref{algebra1})--(\ref{algebra3}). 

\vspace*{.2cm}
Note that the case $n=3$ proceeds in exactly the same way, provided one sets 
$C^{\m}_{~\,\n\r\s}$ to zero and one defines $W^{(3)}_{\O_0} := \tilde{C}_{\a\r\s}\,$. 
In other words, the relations (\ref{algebra1})--(\ref{algebra3}) 
and (\ref{jacobi1})--(\ref{jacobi4}) still hold, setting $C^{\m}_{~\,\n\r\s}=0\,$. 
The representation matrices $\mathbf{\G}^{\a}$ and the Weyl-covariant derivative 
(\ref{Weylderivative}) are unchanged as well.
More explicitly, we have
\begin{eqnarray}
	\underline{n\geqslant 4}&:& \g_1 \cd_{\a_1}C^{\b}_{~\,\g\d\e}
	=\o_{\a}(-\cp_{\m\a_1}^{\a\n}\Delta_{\n}^{~\m})C^{\b}_{~\,\g\d\e}
	\;\leadsto \;\g_1 W_{\O_1} = \o_{\a}\mathbf{\G}^{\a}W_{\O_1}
	\nonumber \\
	\underline{n=3}&:& \g_1 \cd_{\a_1}\tilde{C}_{\g\d\e}
	=\o_{\a}(-\cp_{\m\a_1}^{\a\n}\Delta_{\n}^{~\m})\tilde{C}_{\g\d\e}
	\;\leadsto \;\g_1 W^{(3)}_{\O_1} = \o_{\a}\mathbf{\G}^{\a}_{(3)}W^{(3)}_{\O_1}\,,
	\nonumber
\end{eqnarray}
which shows that the representation matrices $\mathbf{\G}^{\a}$ 
and $\mathbf{\G}_{(3)}^{\a}$ are essentially the same. Indeed, the iterative 
procedure given in the appendix reproduces itself in 
exactly the same way when $n=3$, with the convention that 
$W^{(3)}_{\O_0}\equiv \tilde{C}_{\a\r\s}\,$. 

\vspace*{.3cm}

\section*{Acknowledgements}
The author is grateful to G. Barnich, H. Baum, X. Bekaert, J. Erdmenger, M. Henneaux and 
Ch. Schomblond for stimulating remarks and encouragements. 
F. Brandt is acknowledged for his comments. 
This work was partly done at 
the DAMTP (Cambridge, U.K.), where the author was 
Wiener-Anspach postdoctoral fellow (Belgium). 

\vspace*{.3cm}

\appendix
%******************************************************************
\section{$W$-tensors and their transformations }
\label{A}
%*******************************************************************

The $W$-tensors are computed iteratively, together with their transformation laws under Weyl rescalings of the metric. 

\vspace*{.3cm}

{\bfseries{(A)}} First, we have $\g_1 W_{\O_0} = \o_{\a}\mathbf{\G}^{\a} W_{\O_0} = 0\,$.
Then, we form $W_{\O_1}=\nabla_{\a_1}W_{\O_0}\,$.
Taking the Weyl variation gives
\begin{eqnarray}
	\g_1 W_{\O_1} &=& \g_1 [(\pa_{\a_1}-\G_{\a_1\m}^{~~~~\n}\Delta_{\n}^{~\m})W_{\O_0}]
	              = -\o_{\l}\cp^{\l\n}_{\m\a_1}\Delta_{\n}^{~\m}W_{\O_0}
                \nonumber \\
                &=& \o_{\l}{[T^{\l}]}_{\O_1}^{~\O_0}W_{\O_0}\,,
                \nonumber
\end{eqnarray}
where the last equality serves as a definition for the tensor 
${[T^{\l}]}_{\O_1}^{~\O_0}\,$, which satisfies 
$\g_1 {[T^{\l}]}_{\O_1}^{~\O_0}=0=\nabla_{\m}{[T^{\l}]}_{\O_1}^{~\O_0}\,$. 
We also use the notation $\g_1 W_{\O_1}=\o_{\a}\mathbf{\G}^{\a}W_{\O_0}\,$, cf. equation 
(\ref{defGamma}). 

Continuing, we compute the Weyl variation of $\nabla_{\a_2}W_{\O_1}$:
\begin{eqnarray}
	\g_1 \nabla_{\a_2} W_{\O_1} &=& \nabla_{\a_2}(\o_{\l}{[T^{\l}]}_{\O_1}^{~\O_0}W_{\O_0})
	-\o_{\l}\cp^{\l\n}_{\m\a_2}\Delta_{\n}^{~\m}W_{\O_1}
	\nonumber \\
	&=& (-\g_1 K_{\l\a_2}){[T^{\l}]}_{\O_1}^{~\O_0}W_{\O_0}+
	\o_{\l}{[T^{\l}]}_{\O_1}^{~\O_0}\nabla_{\a_2}W_{\O_0}
  \nonumber \\	
	&&-\,\o_{\l}\cp^{\l\n}_{\m\a_2}\Delta_{\n}^{~\m}W_{\O_1}\,.
	\nonumber
\end{eqnarray}
Using $\g_1({[T^{\l}]}_{\O_1}^{~\O_0}W_{\O_0})=0\,$, we obtain
\begin{eqnarray}
	\g_1 \Big(\nabla_{\a_2} W_{\O_1}&+& K_{\l\a_2}{[T^{\l}]}_{\O_1}^{~\O_0}W_{\O_0}\Big)=
	\nonumber \\
	&&=\,\o_{\l} \Big( \d^{\O'_1}_{\a_2\O_0}{[T^{\l}]}_{\O_1}^{~\O_0} 
	- \d^{\O'_1}_{\O_1}\cp^{\l\n}_{\m\a_2}\Delta_{\n}^{~\m} \Big)W_{\O'_1} 
	\nonumber
\end{eqnarray}
which we rewrite
\begin{eqnarray}
	\g_1 W_{\O_2} = \o_{\l}{[T^{\l}]}_{\O_2}^{~\O_1}W_{\O_1}=\o_{\a}\mathbf{\G}^{\a}W_{\O_2}\,,
	\nonumber
\end{eqnarray}
where $W_{\O_2}\equiv \cd_{\a_2}W_{\O_1}=$
$\nabla_{\a_2} W_{\O_1}+ K_{\l\a_2}{[T^{\l}]}_{\O_1}^{~\O_0}W_{\O_0}\,$.

Calculating $\g_1\g_1W_{\O_2}\,$, we find 
$0 = \o_{\a}\o_{\b}\mathbf{\G}^{\a}\mathbf{\G}^{\b}W_{\O_2}\,$, or 
$[\mathbf{\G}^{\a},\mathbf{\G}^{\b}]=0\,$, cf. second equation of (\ref{algebra1}).
Also, since 
\begin{eqnarray}
W_{\O_2}\equiv \cd_{\a_2}W_{\O_1}\equiv
\cd_{\a_2}\cd_{\a_1}W_{\O_0}=(\nabla_{\a_2}\nabla_{\a_1}+ K_{\l\a_2}{[T^{\l}]}_{\O_1}^{~\O_0})W_{\O_0}\,, 
\nonumber
\end{eqnarray}
we find that 
\begin{eqnarray}
[\cd_{\a_2},\cd_{\a_1}]W_{\O_0}=
C_{\a_2\a_1\m}^{\hspace*{.8cm}\n}\Delta_{\n}^{~\m}W_{\O_0}\,, 
\end{eqnarray}
in agreement with the second equation of (\ref{algebra3}) and 
$\mathbf{\G}^{\a}W_{\O_0}=0$ (equivalent to $\g_1 W_{\O_0}=0$). 

{\bfseries{(B)}} Suppose that we have $W_{\O_k}\equiv\cd_{\a_k}\ldots\cd_{\a_2}\cd_{\a_1}W_{\O_0}\,$, 
$k\geqslant 2\,$. In other words, we know that 
\begin{eqnarray}
W_{\O_k}&=&(\nabla_{\a_k}+K_{\l\a_k}\mathbf{\G}^{\l})W_{\O_{k-1}}\,,   
\nonumber \\
\g_1 W_{\O_k}&=&\o_{\a}\mathbf{\G}^{\a}W_{\O_{k}}
\nonumber \\
&=&\o_{\a}{[T^{\a}]}_{\O_k}^{~\O_{k-1}}W_{\O_{k-1}}\,, 
\nonumber
\end{eqnarray}
and 
\begin{eqnarray}
\mathbf{\G}^{[\a}\mathbf{\G}^{\b]}W_{\O_{k}}=0\,.\nonumber
\end{eqnarray} 
We want to obtain the next tensor, $W_{\O_{k+1}}\equiv\cd_{\a_{k+1}}W_{\O_{k}}$, 
and its transformation rule. 

As before, we first compute the Weyl transformation of $\nabla_{\a_{k+1}}W_{\O_k}\,$:
\begin{eqnarray}
	\g_1 \nabla_{\a_{k+1}}W_{\O_k} &=& \nabla_{\a_{k+1}}\Big(
	\o_{\a}\mathbf{\G}^{\a}W_{\O_{k}}\Big) - 
	\o_{\a}\cp^{\a\m}_{\n\a_{k+1}}\Delta_{\m}^{~\n}W_{\O_k}
	\nonumber \\
	&=&(-\g_1 K_{\a\a_{k+1}})\mathbf{\G}^{\a}W_{\O_{k}}
	+\o_{\a}{[T^{\a}]}_{\O_k}^{~\O_{k-1}}\nabla_{\a_{k+1}}W_{\O_{k-1}}
	\nonumber \\
	&&-\,\o_{\a}\cp^{\a\m}_{\n\a_{k+1}}\Delta_{\m}^{~\n}W_{\O_k}\,.
	\nonumber
\end{eqnarray}
Hence, we get 
\begin{eqnarray}
\g_1\Big(\nabla_{\a_{k+1}}W_{\O_k}&+&K_{\a\a_{k+1}}
\mathbf{\G}^{\a}W_{\O_{k}}\Big) =
K_{\a\a_{k+1}}\o_{\b}\mathbf{\G}^{\a}\mathbf{\G}^{\b}W_{\O_{k}}
\nonumber \\
&&-\,
\o_{\a}\cp^{\a\m}_{\n\a_{k+1}}\Delta_{\m}^{~\n}W_{\O_k}+
\o_{\a}{[T^{\a}]}_{\O_k}^{~\O_{k-1}}\nabla_{\a_{k+1}}W_{\O_{k-1}}\,. 
\nonumber
\end{eqnarray}
Using 
\begin{eqnarray}
\nabla_{\a_{k+1}}W_{\O_{k-1}}=\cd_{\a_{k+1}}W_{\O_{k-1}}-
K_{\b\a_{k+1}}\mathbf{\G}^{\b}W_{\O_{k-1}} \nonumber
\end{eqnarray}
and posing
\begin{eqnarray}
\cd_{\a_{k+1}}W_{\O_k}=\nabla_{\a_{k+1}}W_{\O_k}+K_{\a_{k+1}\l}
\mathbf{\G}^{\l}W_{\O_{k}}\,,\nonumber 
\end{eqnarray}
we find
\begin{eqnarray}	\g_1\cd_{\a_{k+1}}W_{\O_k}&=&K_{\a\a_{k+1}}\o_{\b}\mathbf{\G}^{\a}\mathbf{\G}^{\b}W_{\O_{k}}
	-K_{\b\a_{k+1}}\o_{\a}\mathbf{\G}^{\a}\mathbf{\G}^{\b}W_{\O_{k}}
 \nonumber \\
	&&-\,\o_{\a}\cp^{\a\m}_{\n\a_{k+1}}\Delta_{\m}^{~\n}W_{\O_k}
	+\o_{\a}\d_{\a_{k+1}\O_{k-1}}^{\O'_k}{[T^{\a}]}_{\O_k}^{~\O_{k-1}}W_{\O'_k}
	\nonumber \\
	&=& \o_{\l}\Big(\d_{\a_{k+1}\O_{k-1}}^{\O'_k}{[T^{\a}]}_{\O_k}^{~\O_{k-1}}
	-\d_{\O_{k}}^{\O'_k}\cp^{\l\m}_{\n\a_{k+1}}\Delta_{\m}^{~\n}\Big)W_{\O'_k}
	\nonumber \\
	&=&\o_{\l}{[T^{\a}]}_{\a_{k+1}\O_k}^{~\O'_k}W_{\O'_k}\,,
  \nonumber
\end{eqnarray}
where we used $\mathbf{\G}^{[\a}\mathbf{\G}^{\b]}W_{\O_{k}}=0\,$. $\Box$


\begin{thebibliography}{9}

\bibitem{BE}
N.~Boulanger and J.~Erdmenger, {\it A classification of local Weyl invariants in $D=8$}, 
Class. Quantum Grav. {\bf{21}} (2004) 4305--4316 [{\tt hep-th/0405228}].  

\bibitem{BRST} C.~Becchi, A.~Rouet and R.~Stora,
{\it Renormalization of the abelian Higgs--Kibble model}, Commun. Math. Phys. {\bf 42} (1975) 
127--162; 
{\it Renormalization Of Gauge Theories}, Annals Phys. {\bf 98} (1976) 287--321;
%%CITATION = APNYA,98,287;%%
I.~V.~Tyutin, 
{\it Gauge Invariance In Field Theory And Statistical Physics In Operator
Formalism}, LEBEDEV-75-39; J.~Zinn-Justin, {\textsl{Renormalisation of gauge theories}},
Lecture notes in Physics {n${}^{\circ}$ 37}, Springer, Berlin (1975). 

\bibitem{brandt} F. Brandt, {\it Local BRST Cohomology and Covariance},
Commm. Math. Phys. {\bfseries{190}} (97) 459--489 [{\tt hep-th/9604025}].

\bibitem{jet} F. Brandt, {\it Jet coordinates for local BRST cohomology}, 
Lett. Math. Phys. {\bfseries{55}} (2001) 149--159 [{\tt math-ph/0103006}].
%     SLACcitation  = "%%CITATION = MATH-PH 0103006;%%"

\bibitem{Boulanger:2001he}
N.~Boulanger and M.~Henneaux, {\it A derivation of {W}eyl gravity}, 
Annalen Phys. {\bf 10} (2001) 935--964 [{{\tt hep-th/0106065}}].
%%CITATION = HEP-TH 0106065;%%.

\bibitem{book}
M.~Henneaux and C.~Teitelboim, {\em Quantization of Gauge Systems}, 
Princeton University Press, 1992.

\bibitem{rep} G.~Barnich, F.~Brandt and M.~Henneaux, 
{\it Local {BRST} cohomology in gauge theories}, 
Phys. Rept. {\bf 338} (2000) 439--569 [{{\tt hep-th/0002245}}].
%%CITATION = HEP-TH 0002245;%%.

\end{thebibliography}
\end{document}